
%
%
\normalbaselineskip=12pt
\baselineskip=12pt
\magnification=1200
\hsize 15.0truecm \hoffset 0.0truecm
\vsize 22.0truecm \voffset 0.0truecm
\nopagenumbers
\headline={\ifnum \pageno=1 \hfil \else\hss\tenrm\folio\hss\fi}
\pageno=1

\def\lsim{\mathrel{\rlap{\lower4pt\hbox{\hskip1pt$\sim$}}
    \raise1pt\hbox{$<$}}}         
\def\gsim{\mathrel{\rlap{\lower4pt\hbox{\hskip1pt$\sim$}}
    \raise1pt\hbox{$>$}}}         

\def\ee{e^-e^+}
\def\qq{q\bar q}
\def\rhoqq{\rho_{_{\lambda^{\vphantom{\prime}}_q
 \lambda^{\vphantom{\prime}}_{\bar q};\lambda^{\prime}_q
\lambda^{\prime}_{\bar q}}}(q,\bar q)}
\def\rhoee{\rho_{_{\lambda^{\vphantom{\prime}}_{e^-}
 \lambda^{\vphantom{\prime}}_{e^+};
\lambda^{\prime}_{e^-} \lambda^{\prime}_{e^+}}}(e^-,e^+)}
\def\rhoh{\rho_{_{\lambda^{\vphantom{\prime}}_h \lambda^{\prime}_h}}(h)}
\def\rhoq{\rho_{_{\lambda^{\vphantom{\prime}}_q \lambda^{\prime}_q}}(q)}
\def\fra{D^{\vphantom{*}}_{\lambda^{\vphantom{\prime}}_h
 \lambda^{\vphantom{\prime}}_X; \lambda^{\vphantom{\prime}}_q
 \lambda^{\vphantom{\prime}}_{\bar q}}}
\def\frac{D_{\lambda^{\prime}_h \lambda^{\vphantom{\prime}}_X;
 \lambda^{\prime}_q \lambda^{\prime}_{\bar q}}^*}
\def\La{\Lambda}
\def\la{\lambda}


\line{\hfil DFTT 48/93}
\line{\hfil August 1993}
\line{}
\line{}
\line{}
\centerline{\bf On the parton interpretation of quark fragmentation}
\centerline{\bf into hadrons with different spins}
\vskip 24pt\centerline{ A. Anselm$^{1)}$, M. Anselmino$^{2)}$,
F. Murgia$^{3)}$ and M.G. Ryskin$^{1)}$}
\vskip 12pt
\centerline{1) {\it Petersburg Nuclear Physics Institute, Gatchina,
St. Petersburg, 188350, Russia}}
\vskip 6pt
\centerline{2) {\it Dipartimento di Fisica Teorica, Universit\`a di
Torino and INFN}}
\centerline{\it Via P. Giuria 1, I--10125 Torino, Italy}
\vskip 6pt
\centerline{3) {\it Istituto Nazionale di Fisica Nucleare,
Sezione di Cagliari}}
\centerline{\it Via A. Negri 18, I--09127 Cagliari, Italy}
\vskip 1.2in
\centerline{\bf ABSTRACT}
\vskip 18pt
We calculate the spin density matrix of the hadron $h$ created via quark
fragmentation in the process $e^-e^+ \to q\bar q \to h + X$. In the case
of $h=\Lambda$ the experimental data could possibly elucidate the
problem of $s$-quark contribution to the spin of the $\Lambda$-hyperon,
to be compared with the case of the proton (``spin crisis''). Generally
we find that for hadrons with spin 1/2 the parton description, using
only probabilities, works much better than for vector particles, since
in the former case the non diagonal matrix elements of the hadron spin
density matrix are suppressed by a small factor $p_T/(z\sqrt s)$, where
$p_T$ is the hadron transverse momentum inside the jet.
\vskip 48pt
\vfill
\eject
\baselineskip 12pt plus 1pt minus 1pt
\noindent{\bf 1 - Introduction}
\vskip 12pt
The spin content of baryons, in terms of their constituents, is far from
being clearly understood. The last deep inelastic scattering experiments
[1-3], probing the internal structure of polarized nucleons with
polarized leptons, have upset our original simple spectroscopic picture
of three valence quarks sharing the baryon spin (and other quantum
numbers), while gluons and sea quarks only contribute to the baryon
momentum. Orbital angular momentum, gluon and non perturbative
contributions have also to be invoked and taken into account to
explain the experimental data on polarized nucleon structure functions;
the quarks alone do not seem to be able to account for the proton
spin [4].

Any further experimental information on the internal spin structure of
baryons is therefore of great importance and usefulness in our attempts
of better understanding the subtleties of quark and gluon bound states.
In this respect, as it has been known for some time [5] and recently
reproposed [6,7], the production of $\Lambda$-particles in $e^-e^+$
annihilations seems to be particularly interesting. According to the
$SU(6)$ quark model, the $\Lambda$-hyperon has a very simple
spin-flavour wave function, in that all its spin is carried by the
$s$-quark, while the remaining $ud$ pair is in a $S=I=0$ state.
$s$-quarks in $\ee$ annihilations ($\ee \to s\bar s$) at large energies
are produced strongly polarized; in particular, at the $Z_0$ pole, the
(negative) longitudinal polarization of $s$-quarks can be as high as
$P\simeq -0.9$ and almost independent of the production angle.
Fast $\Lambda$'s produced in $\ee$ annihilations may be thought as the
direct result of $s$-quark fragmentations; moreover, their weak decays
allow a precise measurement of their polarization; it is then natural
to propose a study [6,7] of the correlation between the (observed)
$\Lambda$ polarization and the (computable) $s$-quark polarization to
see to what extent the latter is transferred to the former.

If $s$-quark and $\Lambda$ polarizations turned out to be uncorrelated,
then a ``$\Lambda$ spin crisis'' would add to the ``proton spin crisis''
[8], an intriguing result indeed. Actually, there already exist several
experimental indications that it might be so in the puzzling
polarization data of large $p_T$ $\Lambda$'s (and other hyperons)
inclusively produced in several processes as fragments of unpolarized
nucleons [9]; in all of them it appears that unpolarized
quarks (inside the unpolarized nucleons) give rise to strongly polarized
$\Lambda$'s, what cannot be understood at the constituent level.

In this paper we consider the problem at a more general level by
studying the production of a hadron $h$ in $\ee$ annihilations via
the two-step process $\ee \to q\bar q \to h+X$; we relate the helicity
density matrix of the $h$ hadron, $\rhoh$, to the helicity density
matrix of the $q\bar q$ pair, $\rhoqq$, computed within the Standard
Model, via some ``fragmentation amplitudes'' $\fra (q\bar q \to \La +
X)$. This generalizes the approach of Refs. [6,7] in that it also takes
into account the non diagonal elements of the $\rho(h)$ matrix and
allows for final state interactions among the $q\bar q$ pair; these
final state interactions, normally neglected in the (successful)
computation of jet cross-sections and angular distributions, might be
important when more subtle quantities, like spin correlations, are
involved. Our results reproduce the usual ones [6,7] when only the
independent fragmentation of one quark and diagonal helicity density
matrix elements are considered.

Our approach, a coherent fragmentation picture versus an incoherent
one, was already proposed in a previous paper [10], for the production
of vector mesons in the purely electromagnetic case ({\it e.g.}, $\ee
\to D^*+X$, via one photon annihilation). A related experimental
measurement has also been performed, with the finding of a very small
value of $\rho_{1,-1}(D^*)$ [11]; the large errors, however, do not
allow to draw any definite conclusion and we think that further tests of
the importance of coherence effects, especially in the light of so many
unexpected and subtle spin effects, are important.

For the production of spin 1/2 hadrons it turns out, however, that all
non diagonal spin density matrix elements are strongly suppressed if the
particles inside the jet are well collimated; more precisely, we find
that $\rho_{_{\lambda^{\vphantom{\prime}}_h \lambda^{\prime}_h}}
(h,S_h=1/2) \sim
2p_T/(z\sqrt s)$ for $\lambda^{\vphantom{\prime}}_h
\not= \lambda^{\prime}_h$, where
$p_T$ is the transverse momentum of the hadron with respect to the jet
(original quark) direction and $z\sqrt s/2$ is the hadron longitudinal
momentum. For large energy spin 1/2 hadrons we recover the usual
independent quark fragmentation results which seem to be well founded;
surprisingly, this is not necessarily true for spin 1 vector
particles [10].

In the next Section 2 we discuss our formalism and give the general
expression of the matrix elements $\rhoh$ in our scheme. In Section 3 we
specialize to the case of spin $S =$ 1/2, 1 hadrons; we also recall
how to measure these matrix elements, {\it e.g.} via the angular
distribution of the $\Lambda$ decay products in the $\Lambda$ rest
frame. In the conclusions we discuss our results and the differences
between different spin particles; we stress how clarifying measurements
could be actually performed at LEP or LSC.

\vskip 18pt
\goodbreak

\noindent{\bf 2 - $\ee \to q\bar q \to h+X$ and the helicity density matrix of
particle $h$}
\vskip 12pt
\nobreak

Following Ref. [10] we consider the spin state of a hadron $h$
inclusively produced in $\ee$ annihilation; such a process is supposed
to proceed via the usual two-step process: first a $q\bar q$ pair is
created ($\ee \to q\bar q$) which then annihilates into the observed
hadron plus other unobserved particles ($q\bar q \to h+X$). Whereas the
first process can be perturbatively computed within the Standard Model,
the second one is essentially unknown and parameterized using
phenomenological fragmentation functions. The helicity density matrix of
the hadron $h$ can then be written as
$$\rhoh = {1\over N_h} \sum_{q,X,\lambda^{\vphantom{\prime}}_X,
\lambda^{\vphantom{\prime}}_q,\lambda^{\vphantom{\prime}}_{\bar q},
\lambda^{\prime}_q,\lambda^{\prime}_{\bar q}} \fra \> \rhoqq \> \frac
\eqno(2.1)$$
where $\rhoqq$ is the helicity density matrix of the $q\bar q$ state:
$$\eqalign{
\rhoqq = {1\over N_{q\bar q}} \sum_{\lambda^{\vphantom{\prime}}_{e^-},
\lambda^{\vphantom{\prime}}_{e^+},
\lambda^{\prime}_{e^-},\lambda^{\prime}_{e^+}}
M^{\vphantom{*}}_{\lambda_q \lambda_{\bar q};\lambda_{e^-} \lambda_{e^+}}
&\rhoee \times \cr
&\times M_{\lambda^{\prime}_q \lambda^{\prime}_{\bar q};
\lambda^{\prime}_{e^-} \lambda^{\prime}_{e^+}}^* \cr}\eqno(2.2)$$
The $M$'s are the helicity amplitudes for the $\ee \to \qq$ process and
the $D$'s are the fragmentation {\it amplitudes} for the process $\qq
\to h+X$; they are related to the usual fragmentation {\it functions}
$D^h_q$ by:
$$\sum_{X,\lambda^{\vphantom{\prime}}_X,\lambda^{\vphantom{\prime}}_h,
\lambda^{\vphantom{\prime}}_q,\lambda^{\vphantom{\prime}}_{\bar q},
\lambda^{\prime}_q,\lambda^{\prime}_{\bar q}} \fra \>
\rho_{_{\lambda^{\vphantom{\prime}}_q \lambda^{\vphantom{\prime}}_{\bar q};
\lambda^{\prime}_q \lambda^{\prime}_{\bar q}}} \>
D_{\lambda^{\vphantom{\prime}}_h \lambda^{\vphantom{\prime}}_X;
\lambda^{\prime}_q
\lambda^{\prime}_{\bar q}}^* = D_q^h \eqno(2.3)$$
In Eq. (2.1) a sum is performed over all quark flavours $q$. In
Eqs. (2.1) and (2.3) the $\sum_{X,\lambda_X}$ stays for the phase space
integration and the sum over spins of all the unobserved particles,
grouped into a state $X$. The normalization factors $N_h$ and
$N_{\qq}$ are such that Tr($\rho$) = 1; in particular
$$N_h=
\sum _q D_q^h \eqno(2.4)$$
At last, the spin state of the initial $\ee$ system is described by the
helicity density matrix $\rhoee$. For unpolarized $e^-$ and $e^+$ one
has
$$\rhoee = {1\over 4} \> \delta_{\lambda^{\vphantom{\prime}}_{e^-}
\lambda^{\prime}_{e^-}}
\> \delta_{\lambda^{\vphantom{\prime}}_{e^+}
\lambda^{\prime}_{e^+}} \eqno(2.5)$$

Eq. (2.1) differs from the usual independent quark fragmentation approach
in that the hadronization process is $\qq \to h+X$ rather than $q \to
h+X$; the (necessary) $\qq$ interactions are taken into account in the
fragmentation amplitudes. Indeed, if one assumes the $\fra$ to be
independent of $\bar q$ and ignores the quantum numbers of $\bar q$ and
all its fragmentation products (so that one can set
$\lambda^{\vphantom{\prime}}_{\bar q} =
\lambda^{\prime}_{\bar q}$ in $\rho(q,\bar q)$ and uses
$\sum_{\lambda^{\vphantom{\prime}}_{\bar q}}
\rho_{_{\lambda^{\vphantom{\prime}}_q\lambda^{\vphantom{\prime}}_{\bar q};
\lambda^{\prime}_q\lambda^{\vphantom{\prime}}_{\bar q}}} =
\rho_{_{\lambda^{\vphantom{\prime}}_q\lambda^{\prime}_q}}(q)$),
Eq. (2.1) gives
$$\rhoh = {1\over N_h} \sum_{q,X,\la^{\vphantom{\prime}}_X,
\la^{\vphantom{\prime}}_q,\la^{\prime}_q}
D^{\vphantom{*}}_{\la^{\vphantom{\prime}}_h
\la^{\vphantom{\prime}}_X;\la^{\vphantom{\prime}}_q} \> \rhoq \>
D^*_{\la^{\prime}_h\la^{\vphantom{\prime}}_X;\la^{\prime}_q} \eqno(2.6)$$
Moreover, if the final hadron momentum is parallel to the quark one,
angular momentum conservation in the forward direction requires, for
each $D_{\la_h\la_X;\la_q} (q \to h+X)$,
$$ \la_h + \la_X = \la_q  \eqno(2.7)$$
Remembering also that, when neglecting quark masses, the helicity
density matrix of a quark produced in $\ee \to \qq$ annihilation is
diagonal ($\rhoq = 0$ for $\lambda^{\vphantom{\prime}}_q
\not= \lambda^{\prime}_q$) we have
that only the diagonal terms of Eq. (2.6) survive, giving the usual
probabilistic formula [5]
$$\rho_{_{\lambda_h\lambda_h}}(h) = {1\over N_h} \sum_{q,\lambda_q}
\rho_{_{\lambda_q\lambda_q}}(q) D_{q,\lambda_q}^{h,\lambda_h} \eqno(2.8)$$
where $D_{q,\lambda_q}^{h,\lambda_h} = \sum_{X,\lambda_X} \vert
D_{\lambda_h,\lambda_X;\lambda_q} \vert^2$.

This needs not be true in general [10], as it can be seen from
Eqs. (2.1,2). The $\qq$ helicity density matrix can be computed from the
knowledge of $\rho(e^-,e^+)$ (Eq. (2.5), for unpolarized beams) and the
center of mass annihilation amplitudes $M_{\lambda_q\lambda_{\bar q};
\lambda_{e^-}\lambda_{e^+}}$. At lowest order in the perturbative
Standard Model, neglecting quark masses and including both the weak
($Z_0$) and the electromagnetic ($\gamma$) contributions, they are
given, for unpolarized initial leptons ($l$) and in terms of the usual
Standard Model parameters [12], by
$$ \eqalign{&M_{\lambda_q\lambda_{\bar q};\lambda_{l^-}\lambda_{l^+}}
(s,\theta)= i\delta_{\lambda_{l^-},-\lambda_{l^+}} \>
\delta_{\lambda_q,-\lambda_{\bar q}} \times \cr
&\times \Big\{ \Big[ e_q e^2-{1\over 4}
g_Z^2 c_V^l c_V^q {s\over s-M_Z^2 +iM_Z\Gamma_Z}\Big]
(1+4\lambda_{l^-}\lambda_q \cos\theta) \cr
& +{1\over 4}g_Z^2 {s\over s-M_Z^2 +iM_Z\Gamma_Z} \big[
2c_V^lc_A^q(\lambda_{l^-}\cos\theta + \lambda_q) \cr
&+ 2c_A^lc_V^q
(\lambda_{l^-}+\lambda_q\cos\theta) - c_A^lc_A^q(\cos\theta +
4\lambda_{l^-}\lambda_q) \big] \Big\} \cr} \eqno(2.9)$$
where $\sqrt s$ is the $l^-l^+$ c.m. energy, $\theta$ the $q$ production
angle and $e_q$ the quark charge. From Eqs. (2.2), (2.5) and (2.9) it can be
seen that the only non zero elements of $\rhoqq$ are
$$\eqalign{
\rho_{+-;+-}(\qq) =& 1-\rho_{-+;-+}(\qq) \cr
\rho_{+-;-+}(\qq) =& \rho_{-+;+-}^*(\qq) \cr} \eqno(2.10)$$

Inserting Eq. (2.10) into Eq. (2.1) obtains
$$\eqalign{ \rhoh = {1\over N_h}&\sum_{X,\la^{\vphantom{\prime}}_X}\left\{
\left[ D^{\vphantom{*}}_{\la^{\vphantom{\prime}}_h\la^{\vphantom{\prime}}_X;+-}
D^*_{\la^{\prime}_h\la^{\vphantom{\prime}}_X;+-}
      -D^{\vphantom{*}}_{\la^{\vphantom{\prime}}_h\la^{\vphantom{\prime}}_X;-+}
D^*_{\la^{\prime}_h\la^{\vphantom{\prime}}_X;-+}
\right] \rho_{+-;+-} \right. \cr
&+ \left[ D^{\vphantom{*}}_{\la^{\vphantom{\prime}}_h
\la^{\vphantom{\prime}}_X;+-}
D^*_{\la^{\prime}_h\la^{\vphantom{\prime}}_X;-+}
         +D^{\vphantom{*}}_{\la^{\vphantom{\prime}}_h
\la^{\vphantom{\prime}}_X;-+} D^*_{\la^{\prime}_h\la^{\vphantom{\prime}}_X;+-}
\right] {\rm Re}[\rho_{+-;-+}] \cr
&+i\left[ D^{\vphantom{*}}_{\la^{\vphantom{\prime}}_h
\la^{\vphantom{\prime}}_X;+-} D^*_{\la^{\prime}_h\la^{\vphantom{\prime}}_X;-+}
         -D^{\vphantom{*}}_{\la^{\vphantom{\prime}}_h
\la^{\vphantom{\prime}}_X;-+} D^*_{\la^{\prime}_h\la^{\vphantom{\prime}}_X;+-}
\right] {\rm Im}[\rho_{+-;-+}] \cr
&\left.+D^{\vphantom{*}}_{\la^{\vphantom{\prime}}_h
\la^{\vphantom{\prime}}_X;-+} D^*_{\la^{\prime}_h\la^{\vphantom{\prime}}_X;-+}
\right\}\cr}\eqno(2.11)$$

The above expression of $\rhoh$ can be further simplified by exploiting
the fact that the hadronization process ($\qq \to h+X$) is parity
invariant; the parity relations for c.m. helicity amplitudes [13] then
yield
$$\sum_{\la^{\vphantom{\prime}}_X} \fra\frac =
(-1)^{2S^{\vphantom{\prime}}_h-\la^{\vphantom{\prime}}_h-\la^{\prime}_h}
   D^{\vphantom{*}}_{-\la^{\vphantom{\prime}}_h
\la^{\vphantom{\prime}}_X;-\la^{\vphantom{\prime}}_q-
\la^{\vphantom{\prime}}_{\bar q}}
   D^*_{-\la^{\prime}_h \la^{\vphantom{\prime}}_X;-\la^{\prime}_q
       -\la^{\prime}_{\bar q}} \eqno(2.12)$$

It appears from Eq. (2.11) that the non diagonal elements of
$\rho(\qq)$, usually neglected in the independent quark fragmentation
scheme, contribute both to
$\rho_{_{\la^{\vphantom{\prime}}_h \la^{\vphantom{\prime}}_h}}(h)$
and $\rho_{_{\la^{\vphantom{\prime}}_h \la^{\prime}_h}}(h)
\> (\la^{\vphantom{\prime}}_h \not= \la^{\prime}_h)$; in particular
the latter can be different from zero. However, even if the
fragmentation amplitudes $\fra$ are essentially unknown, we know from
experiment that the hadron production, at lowest perturbative order in
$\ee$ annihilation, proceeds via the creation of two collimated jets of
particles, each of which retains the original $q$ and $\bar q$
direction. The $\fra$ can then be regarded as center of mass helicity
amplitudes for the process $\qq \to h+X$, {\it essentially in the
forward direction}, that is with the hadron momentum $\vec h$ almost
parallel to the quark one , $\vec q$ (and $\vec {\bar q}$ parallel to
$\vec X$).

{}From the well known forward behaviour of c.m. helicity amplitudes [13]
then we have
%
%
$$\fra\frac \sim
\left( \sin{\theta_h\over 2}
\right)^{\vert \la^{\vphantom{\prime}}_h-
\la^{\vphantom{\prime}}_X-\la^{\vphantom{\prime}}_q+
\la^{\vphantom{\prime}}_{\bar q}\vert
+\vert \la^{\prime}_h-\la^{\vphantom{\prime}}_X-
\la^{\prime}_q+\la^{\prime}_{\bar q}\vert}
\eqno(2.13)$$
where $\theta_h$ is the angle between the hadron momentum, $\vec h = z
\vec q + \vec p_T$, and the quark momentum $\vec q$, that is
$$ \sin\theta_h = {2p_T \over z\sqrt s}, \eqno(2.14)$$
where we have used $\vert \vec q \vert = \sqrt s/2$.

The bilinear combinations of fragmentation amplitudes contributing to
$\rho(h)$ are then not suppressed by powers of $(p_T/(z\sqrt s))$ only
if the exponents in Eq. (2.13) are zero; which entails
$$\la^{\vphantom{\prime}}_h - \la^{\prime}_h =
(\la^{\vphantom{\prime}}_q - \la^{\vphantom{\prime}}_{\bar q}) -
(\la^{\prime}_q - \la^{\prime}_{\bar q}). \eqno(2.15)$$

Eqs. (2.11-14) hold in general, for the production of any hadron $h$
with spin $S_h$. In the next Section we specialize them to the cases
$S_h = 1/2$ and $S_h = 1$.

\vskip 18pt
\goodbreak

\noindent{\bf 3 - $\ee \to \qq \to h+X, \> S_h = 1/2, 1$}
\vskip 12pt
\nobreak

{}From Eqs. (2.11) and (2.12) we have, for the helicity density matrix of
hadrons with spin $S_h = 1/2$
$$\eqalignno{&\rho_{++}(S_h=1/2) = {1\over N_h}\sum_{X,\la_X,q} \left\{
\vert D_{+\la_X;-+} \vert^2 + \left[  \vert D_{+\la_X;+-} \vert^2
- \vert D_{+\la_X;-+} \vert^2 \right] \rho_{+-;+-} \right. \cr
&\quad\quad\quad + 2{\rm Re} \left[ D_{+\la_X;+-}D^*_{+\la_X;-+}
\right] {\rm Re} [\rho_{+-;-+}] \cr
&\quad\quad\quad \left. - 2{\rm Im} \left[ D_{+\la_X;+-}D^*_{+\la_X;-+}
\right] {\rm Im} [\rho_{+-;-+}] \right\} &(3.1) \cr
&{\rm Re}[\rho_{+-}(S_h=1/2)] = {1\over N_h}\sum_{X,\la_X,q} \left\{
{\rm Re}\left[ D_{+\la_X;+-}D^*_{-\la_X;+-}\right]
(2\rho_{+-;+-}-1) \right. \cr
&\quad\quad\quad \left. - {\rm Im}\left[ D_{+\la_X;+-}D^*_{-\la_X;-+}
- D_{+\la_X;-+}D^*_{-\la_X;+-} \right]
{\rm Im} [\rho_{+-;-+}] \right \} &(3.2) \cr
&{\rm Im}[\rho_{+-}(S_h=1/2)] = {1\over N_h}\sum_{X,\la_X,q} \left\{
{\rm Im}\left[ D_{+\la_X;+-}D^*_{-\la_X;+-}\right] \right. \cr
&\quad\quad\quad \left. + {\rm Im}\left[ D_{+\la_X;+-}D^*_{-\la_X;-+}
+ D_{+\la_X;-+}D^*_{-\la_X;+-} \right]
{\rm Re} [\rho_{+-;-+}] \right \} &(3.3) \cr}$$

However, in this case, none of the non diagonal bilinear combinations
appearing in the above equations satisfy Eq. (2.15). That is, in the
$p_T/(z\sqrt s) \to 0$ limit, we recover the usual probabilistic
independent quark fragmentation results (2.8)
$$\eqalign{ \rho_{++}(S_h=1/2)&={1\over N_h}\sum_q \left[
\rho_{++}(q) D_{q,+}^{h,+} + \rho_{--}(q) D_{q,-}^{h,+} \right] \cr
\rho_{+-}(S_h=1/2)&=0 \cr} \eqno(3.4)$$
where we have used $\rho_{+-;+-}(\qq) = \rho_{++}(q)$. Notice that the
corrections to Eq. (3.4) are of order $(p_T/(z\sqrt s))^2$ for
$\rho_{++}$ and $(p_T/(z\sqrt s))$ for $\rho_{+-}$.

The same conclusion does not hold for the production of vector
particles, $S_h = 1$ [10]. In such case Eq. (2.15) can be satisfied by
$\la^{\vphantom{\prime}}_h = -\la^{\prime}_h = 1, \>
\la^{\vphantom{\prime}}_q-\la^{\vphantom{\prime}}_{\bar q} =
\la^{\prime}_{\bar q} - \la^{\prime}_q = 1$ and, even in the
$(p_T/(z\sqrt s)) \to 0$ limit, one remains with non zero non diagonal
matrix elements:
$$\eqalign{
{\rm Re}[\rho_{1,-1}(S_h=1)] &= {1\over N_h} \sum_{X,\la_X,q}
D_{1\la_X;+-}D^*_{-1\la_X;-+} {\rm Re}[\rho_{+-;-+}] \cr
{\rm Im}[\rho_{1,-1}(S_h=1)] &= {1\over N_h} \sum_{X,\la_X,q}
D_{1\la_X;+-}D^*_{-1\la_X;-+} {\rm Im}[\rho_{+-;-+}] \cr}\eqno(3.5)$$

The helicity density matrix elements can be measured by looking at the
two-body decay of the hadron $h$ in its helicity rest frame,
$h \to A+B$.
Let us consider as two most common examples the decays of a spin 1/2
hadron into a spin 1/2 + a spin 0 ({\it e.g.} $\La \to p\pi^-, \Sigma^+
\to p\pi^0, \La_c \to \La \pi^+$) and of a spin 1 into two spin 0
({\it e.g.} $\rho \to \pi\pi, K^* \to K\pi, D^* \to D\pi$). In the
former case the normalized angular distribution of the final particle
$A$ ({\it e.g.} $p$, for $\La \to p\pi$ decay) is given by
$$\eqalign{ W(\theta_A, \varphi_A) &= {1\over 2\pi} \Big\{
{1\over 2} - {1\over 2} \alpha \cos\theta_A + \alpha \rho_{++}(S_h=1/2)
\cos\theta_A  \cr
&+ \alpha {\rm Re}[\rho_{+-}(S_h=1/2)] \sin\theta_A \cos\varphi_A \cr
&- \alpha {\rm Im}[\rho_{+-}(S_h=1/2)]
\sin\theta_A \sin\varphi_A \Big\} \cr} \eqno(3.6)$$
where $\theta_A$ and $\varphi_A$ are respectively the polar and
azimuthal angle of particle $A$ in the rest frame of the decaying
spin 1/2 hadron; $\alpha$ is the known weak decay parameter ({\it e.g.},
$\alpha = 0.642 \pm 0.013$ for $\La \to p\pi^-$ decay).

In the case of a spin 1 $\to$ spin 0 + spin 0 decay one has
$$\eqalign{ W(\theta_A, \varphi_A) &= {3\over 4\pi} \Big\{
{1\over 2}(1-\rho_{0,0}) + {1\over 2}(3\rho_{0,0}-1)
\cos^2\theta_A \cr
&-{1\over \sqrt 2} \sin 2\theta_A \cos\varphi_A {\rm Re}
[\rho_{1,0} - \rho_{0,-1}] \cr
&+{1\over \sqrt 2} \sin 2\theta_A \sin\varphi_A {\rm Im}
[\rho_{1,0} - \rho_{0,-1}] \cr
&-\sin^2\theta_A \cos 2\varphi_A {\rm Re}[\rho_{1,-1}]
+\sin^2\theta_A \sin 2\varphi_A {\rm Im}[\rho_{1,-1}]\Big\}
\cr} \eqno(3.7)$$

In the case in which the production process of the hadron $h$ is parity
invariant ($\ee \to h + X$ at $ \sqrt s \ll M_Z$, when weak effects can
be neglected) we have further parity relations between the matrix
elements of $\rho(h)$. Eqs. (3.6) and (3.7) then simplify respectively
to
$$W(\theta_A, \varphi_A) = {1\over 2\pi} \left\{
{1\over 2} - \alpha {\rm Im}[\rho_{+-}] \sin\theta_A \sin\varphi_A
\right\} \eqno(3.8)$$
\line{}
$$\eqalign{ W(\theta_A, \varphi_A) &= {3\over 4\pi} \Big\{
{1\over 2}(1-\rho_{0,0}) + {1\over 2}(3\rho_{0,0}-1)
\cos^2\theta_A \cr
&-\sqrt 2 \sin 2\theta_A \cos\varphi_A {\rm Re} [\rho_{1,0}]
-\sin^2\theta_A \cos 2\varphi_A \rho_{1,-1} \Big\} \cr} \eqno(3.9)$$

A measurement of the spin density matrix elements of $D^*$
mesons produced in $\ee$ annihilation at $\sqrt s = 29$ GeV, according
to Eq. (3.9), has been performed [11]. The data seem to be consistent
with zero values of $\rho_{_{\la\la^{\prime}}}(D^*) \> (\la \not=
\la^{\prime})$, but the large errors do not allow yet any definite
conclusion.

\vskip 18pt
\goodbreak

\noindent{\bf 4 - Conclusions}
\vskip 12pt
\nobreak

The study of spin properties of hadrons produced in $\ee$ annihilations
should supply a rich and interesting ground for understanding the
relationship between the spin of the constituents and that of the
composite mesons and baryons. Such understanding is at the moment, after
the so called ``proton spin crisis'' triggered by DIS experiments,
rather confused and in need of much better theoretical and experimental
information.

In the large energy $\ee$ experiments which are being performed both at
LEP and SLC, quarks are always produced, via electro-weak interactions,
with a large, well known polarization; to what extent this is
transferred to the final observed hadrons is an open question which
should be addressed and hopefully answered in the near future. The usual
approach is a simple probabilistic scheme, according to which a
polarized quark fragments independently into the final hadron, and our
ignorance on the hadronization process is hidden in phenomenological
fragmentation functions which should be measured via inclusive $\ee \to
h+X$ cross-sections.

Such a scheme is rather successful when measuring jet angular
distributions and unpolarized cross-sections, in that the amount and
directions of the original quarks reflect accurately into the amount and
directions of the observed hadrons. This might not be true when more
subtle quantities, like spin observables, are involved; indeed, many
mysterious spin effects in several physical processes prove to be
a challenge for the existing theories. The reason is that spin variables
often involve subtle interference effects between different amplitudes
rather than simple squared moduli, like unpolarized cross-sections.

In this paper we have considered a more general scheme, namely $\ee
\to \qq$ and $\qq \to h+X$, as a two-step process to describe hadron
production in $\ee$ annihilations. The second part, $\qq \to h+X$, is
treated as an effective two-body into two-body process which takes
into account the $\qq$ interactions and the full spin states of the
initial $\qq$ system; these can be easily computed within the
perturbative Standard Model.

Our general result for the spin density matrix of the hadron $h$ is
given by Eq. (2.11) and it differs from the usual independent quark
fragmentation; for example, it allows non zero non diagonal matrix
elements. However, when exploiting further experimental information,
like the narrowness of the observed jets, it turns out that many of the
contributions to Eq. (2.11) are small and negligible at large energies.
Surprisingly, whether one recovers the usual probabilistic results or
not, depends on the spin of the final hadron $h$: for spin 1/2 particles
this is indeed the case, whereas for spin 1 vector particles one can
still remain with non diagonal matrix elements. Only a direct
measurement of these matrix elements would definitely settle the
question.

 The fact that for spin 1/2 hadrons the independent fragmentation of one
quark turns out to be a well founded model is an encouraging information
in view of the proposed analysis of the $\La$-hyperon polarization: to a
good approximation a fast $\La$ produced in $\ee$ annihilation is
generated by a strongly polarized $s$-quark. Its independent
fragmentation into the $\La$ should certainly determine the spin of the
latter; if entirely, that is with the $s$-quark and the $\La$
polarization equal, or only partially, is something we do not know and
we have to investigate. The answer depends on the details and
subtleties of the $\La$ wave function and might not be trivial, as the
proton spin crisis has taught us; having reached the conclusion that
$\qq$ interactions should not affect the analysis is already a first
reassuring step in the difficult and hard task in front of us.

\vskip 24pt
\goodbreak

\noindent{\bf Acknowledgements}
\vskip 6pt
\nobreak

One of us (M.A.) would like to thank the Petersburg Nuclear Physics
Institute, where part of this work was done, for the warm and kind
hospitality.

\vskip 36pt
\goodbreak

\noindent{\bf References}
\vskip 12pt
\nobreak

\item {[ 1]} J.~Ashman {\it et al.}, {\it Phys. Lett.} {\bf B206} (1988)
364; {\it Nucl. Phys.} {\bf B328} (1989) 1
\item{[ 2]} B.~Adeva {\it et al.}, {\it Phys. Lett.} {\bf B302} (1993) 533
\item{[ 3]} R.~Arnold {\it et al.}, {\it SLAC preprint} SLAC PUB 5101 (1993)
\item{[ 4]} For reviews see , {\it e.g.}, G.~Altarelli, in ``The
Challenging Questions'', {\it Proceedings} of the 1989 Erice School,
A. Zichichi Ed. (Plenum, New York, 1990); J.~Ellis, {\it Nucl. Phys.}
{\bf A546} (1992) 447c
\item{[ 5]} See, {\it e.g.}, N.S.~Craigie, H.~Hidaka, M.~Jacob
and F.M.~Renard, {\it Phys. Rep.} {\bf 99} (1983) 69
\item{[ 6]} G.~Gustafson and J.~H\"akkinen, {\it Lund Preprint} LU TP
92-29 (1992)
\item{[ 7]} M.~Burkardt and R.L.~Jaffe, {\it MIT preprint} CPT\#2186 (1993)
\item{[ 8]} M.~Anselmino and E.~Leader, {\it Z. Phys.} {\bf C41} (1988)
239
\item{[ 9]} See, {\it e.g.}, K.~Heller, in {\it Spin and Polarization
Dynamics in Nuclear \& Particle Physics},
Eds. A.O.~Barut, Y.~Onel and A.~Penzo
(World Scientific, Singapore, 1990)
\item{[10]} M.~Anselmino, P.~Kroll and B.~Pire, {\it Z. Phys.} {\bf
C29} (1985) 135
\item{[11]} S.~Abachi {\it et al.}, {\it Phys. Lett.} {\bf B199} (1987)
585
\item{[12]} See, {\it e.g.}, D. Griffiths, ``Introduction to
Elementary Particles'', (Harper \& Row, New York, 1987)
\item{[13]} See, {\it e.g.}, C.~Bourrely, E.~Leader and J.~Soffer, {\it
Phys. Rep.} {\bf 59} (1980) 95

\bye